\documentclass[prl,nobibnotes,floatfix,superscriptaddress,twocolumn,showpacs]{revtex4-1}
\usepackage{amsfonts,amsmath,graphicx,epsfig,latexsym}
\usepackage{subfigure}
\usepackage{graphicx}
\usepackage{color}
\usepackage{natbib}
\usepackage{multirow}
\usepackage{latexsym,amssymb}
\usepackage{amsmath}
\begin{document}
\title{ Strong Pressure Dependent Electron-Phonon Coupling in FeSe}

\author{Subhasish Mandal}
\affiliation
{Geophysical Laboratory, Carnegie Institution of Washington, Washington D.C. 20015, USA}
\author{R. E. Cohen}
\affiliation
{Geophysical Laboratory, Carnegie Institution of Washington, Washington D.C. 20015, USA }
\affiliation
{Department of Earth Sciences, University College London, Gower Street, WC1E 6BT, London, United Kingdom}
\author{K. Haule}
\affiliation
{Department of Physics, Rutgers University, Piscataway, New Jersey 08854, USA}

\begin{abstract}
{\footnotesize
We have computed the correlated electronic structure of FeSe and its dependence on the $A_{1g}$ mode versus compression. Using the self-consistent density functional theory - dynamical mean field theory (DFT-DMFT)
 with continuous time quantum Monte Carlo (CTQMC), we find that there is greatly enhanced 
 coupling between some correlated electron states and the $A_{1g}$ lattice distortion. Superconductivity in FeSe shows a very strong sensitivity to pressure, 
with an increase in T$_c$ of almost a factor of 5 within a few GPa, followed by a drop, despite monotonic pressure dependence of almost all electronic properties. 
We find that the maximum $A_{1g}$ deformation potential behaves similar to the experimental $T_c$. In contrast, the maximum deformation potential in DFT for this mode 
increases monotonically with increasing pressure. }
\end{abstract}

\pacs{74.70.Xa, 74.25.Jb, 75.10.Lp} 

\maketitle
\newpage

So far there is no predictive theory for superconductivity in the cuprate and iron-superconductors, hence these superconductors can be classified as non-conventional superconductors, since there is a well-developed, predictive theory for electron-phonon superconductors whose normal state is well-represented by conventional density functional theory (DFT) \cite{PhysRevLett.108.045502,PhysRevB.74.094519,PhysRevLett.64.2575}.  The unconventional superconductors are very sensitive to applied pressure, so pressure provides a control to test theories and develop a better understanding \cite{naturemat,prl1,nature1}. Here we study superconductivity under applied pressure in pure FeSe; unlike cuprates, superconductivity in 
FeSe arises without doping. FeSe is an ideal system to study the electron pairing mechanism due to the simplicity of its crystal and electronic structure. It shows a very strong enhancement of $T_c$ upon application of modest pressure with dramatic increase of $T_c$ from
 8K to $\sim$ 37K\cite{naturemat,prb1,prl2} and then decreases upon further application of pressure.
% specially for the low pressure range (0-3 GPa)  $T_c$ was found to increase rapidly and can reach up to 27K at 1.48 GPa\cite{Mizuguchi:2008bn}. 
Why does $T_c$ increase with pressure and then decrease for rather small lattice compression? This question was addressed in Ref. \onlinecite{prl2}, where it was found that applied pressure (P) intensified antiferromagnetic spin fluctuations (SF). However, this did not explain the decrease in $T_c$ with further compression. 

The discovery of the iron superconductors showed that high T$_c$ is not specific to the cuprates, and suggests a wider field of potential high T$_c$ materials \cite{Mazin:2010he}. Although DFT gives many properties reasonably accurately for both cuprates \cite{Pickett:1990uh,Cohen:1989vj,Pickett:1992} and Fe-superconductors\cite{Subedi:2008hc,Johannes:2010iz,Mazin:2008gu}, there is also significant indication of the importance of correlations and fluctuating local moments beyond DFT specially for the Fe-superconductors which are paramagnetic metals in room temperature, and Dynamical Mean Field Theory (DMFT) has proved to be a good approximation \cite{RevModPhys.78.865,kotliar:53,Schafgans:2012kl,dmft2,Yin:2011ca,haule3,Kunes:2008bh}.

 While most studies suggest a spin fluctuation coupling mechanism for SC in Fe-superconductors similarly to cuprates\cite{PhysRevLett.101.057003,PhysRevLett.101.057008,SF3}, strong coupling phonons have also been proposed to play a role in both sets of materials \cite{Anisimov:1992,Cohen:1994,Boeri:2010un,Alexandrov:1994wd,PhysRevLett.64.2575,Chen:2007,*Dewaele:2009ur}. 
The study of strong electron-phonon coupling in correlated solids is in its infancy due to extreme computational complexity. Only in the non-self-consistent Hubbard I and LDA+U approximations has it proved  tractable so far \cite{Savrasov:2003dy,*Floris:2011ca}.  
The role of lattice vibrations in the mechanism of SC in unconventional superconductors is still controversial. The observation of strong electron-phonon and spin-phonon coupling, both in cuprates 
\cite{ep1,HKrakauer:2011vv,Anonymous:YZr5HBL9,ep2} and iron superconductors\cite{isotop2,ep3,ep4,PhysRevB.83.245127,Kumar2010557,Liu:2013vb} indicates that the electron-phonon coupling (EPC) may play an important role in the unconventional superconductors, 
at least to explain the observed Fe-isotope effect\cite{Khasanov:2010bu}, the anomalous temperature dependence of the local Fe-As displacement\cite{Higashitaniguchi:2008id}, gap anisotropy, and the correlation of $T_c$ with the Fe-anion height \cite{Okabe:2010bw}.
These observations also suggest polaron and/or bipolaron driven superconductivity in this material \cite{BussmannHolder:2008fh,BussmannHolder:2009fw,Zhao:1997wv,Salje:2005ts,alex,BussmannHolder:2008cl}.

Here we examine the effects of pressure on the electronic structure and electron-phonon interaction (EPI) in FeSe 
using DMFT in combination with the DFT as implemented in \cite{PhysRevB.81.195107}.

\begin{figure}
\includegraphics[width=220pt, angle=0]{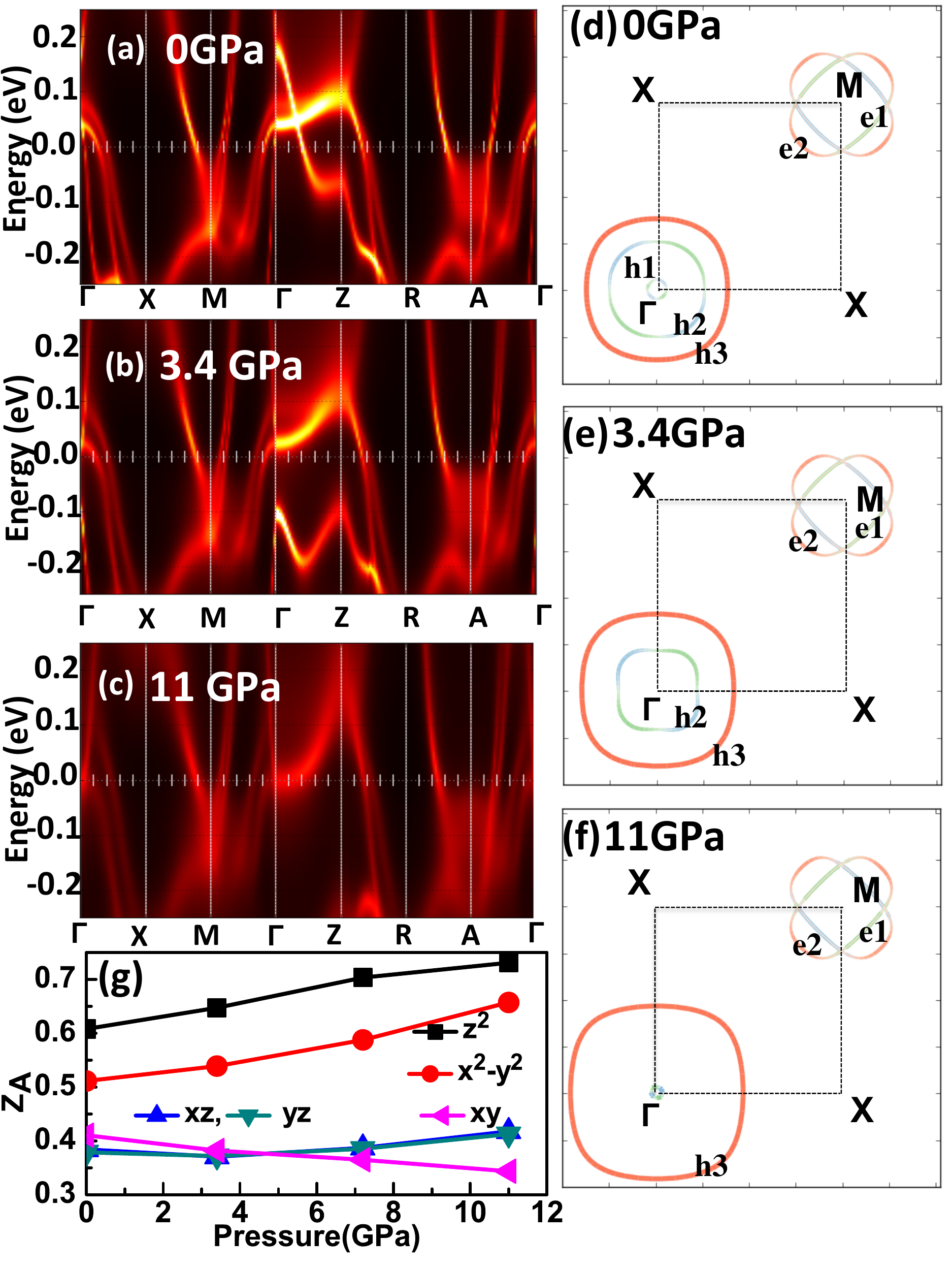}
\caption{ (Color online)
Pressure evolution of DFT-DMFT spectral function (left panel, a-c) and the Fermi surface on the $k_z$=0 plane (right panel, d-f). 
 The Fermi surface is colored in red, green, and blue according to its dominating orbital character of xy, xz, and yz respectively.
(g) Quasiparticle weight as a function of pressure for the Fe-$d$ orbitals.   
 }
\end{figure}

%\twocolumn

Electronic and magnetic properties of FeSe are very sensitive to the position ($z_{Se}$) 
of the selenium layers with respect to the iron layers\cite{Moon:2010fr,Okabe:2010bw,z3,Kuroki:2009bn,Yin:2008id,Essenberger:2012gh,Mazin:2008gu}. The magnetic ordering is found to be strongly affected  by the chalcogen height\cite{Moon:2010fr}.
An accurate estimation of $z_{Se}$ is essential to study the electronic structure of FeSe. 
We optimized $z_{Se}$  for P=0, 3.4, 7.2, and 11 GPa and notice a monotonic increase with P.
The reported experimental values of $z_{Se}$ at ambient pressure are consistently estimated to $z_{Se}$ = 0.267\cite{PhysRevB.79.014519}.
Our DFT+DMFT computations give value of $z_{Se}$=0.27 at ambient pressure (Fig. S2 in Supplementary Information), 
whereas LDA and spin-polarized GGA give 0.234\cite{Subedi:2008hc} and 0.26 respectively. The inclusion of local spin fluctuations by DMFT is hence crucial to describe the structural 
properties in the disordered paramagnetic state, and has similar effect on $z_{Se}$ as the presence of long ranged order in standard DFT.

We compute the DFT-DMFT spectral function ($A(\omega,k) $) and the Fermi surface (FS) 
for the optimized value of $z_{Se}$ at different pressures and summarized them in the Fig. 1. 
A dramatic change in  $A(\omega,k) $  is noticed when the pressure is increased from 0 to 3.4 GPa. There are three DFT-DMFT
bands crossing the Fermi energy ($E_F$) from $\Gamma$ to $X$-point at P=0, while there are only two bands crossing $E_F$ at P=3.4 GPa. The strong 3D band which crosses $E_F$ from $\Gamma$ to $Z$ at P=0,
does not cross  $E_F$ above 3.4 GPa. Further, we consider the orbital resolved FS at $k_z$=0 plane as a function of pressure. From Fig. 1(d-f) we notice that the electron pockets (e1 and e2) at the M point do not change much with increasing the
pressure whereas the hole pockets (h1, h2, and h3) at the $\Gamma$ point changes significantly. The number of hole pocket􀀀 reduces from three to one by increasing the pressure. The outer
hole pocket h3 are mainly of xy character, while the inner hole pockets h1 and h2 consists of xz and yz character.  
With increasing pressure three hole pockets behave differently. The outer hole pocket h3 expands, while the inner hole pocket h2 shrinks with increasing pressure.
The inner hole pocket h1, which shows  most $k_z$ dependence, vanishes above P=3.4 GPa.

\begin{figure}
\includegraphics[width=220pt, angle=0]{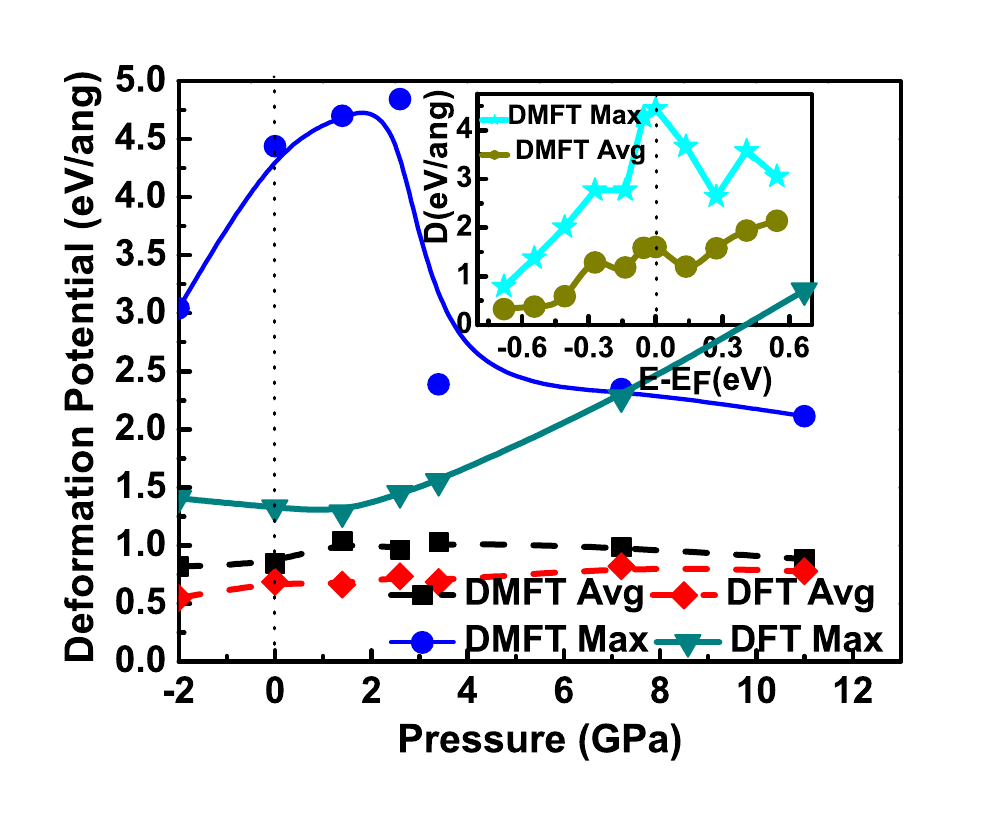}
\includegraphics[width=220pt, angle=0]{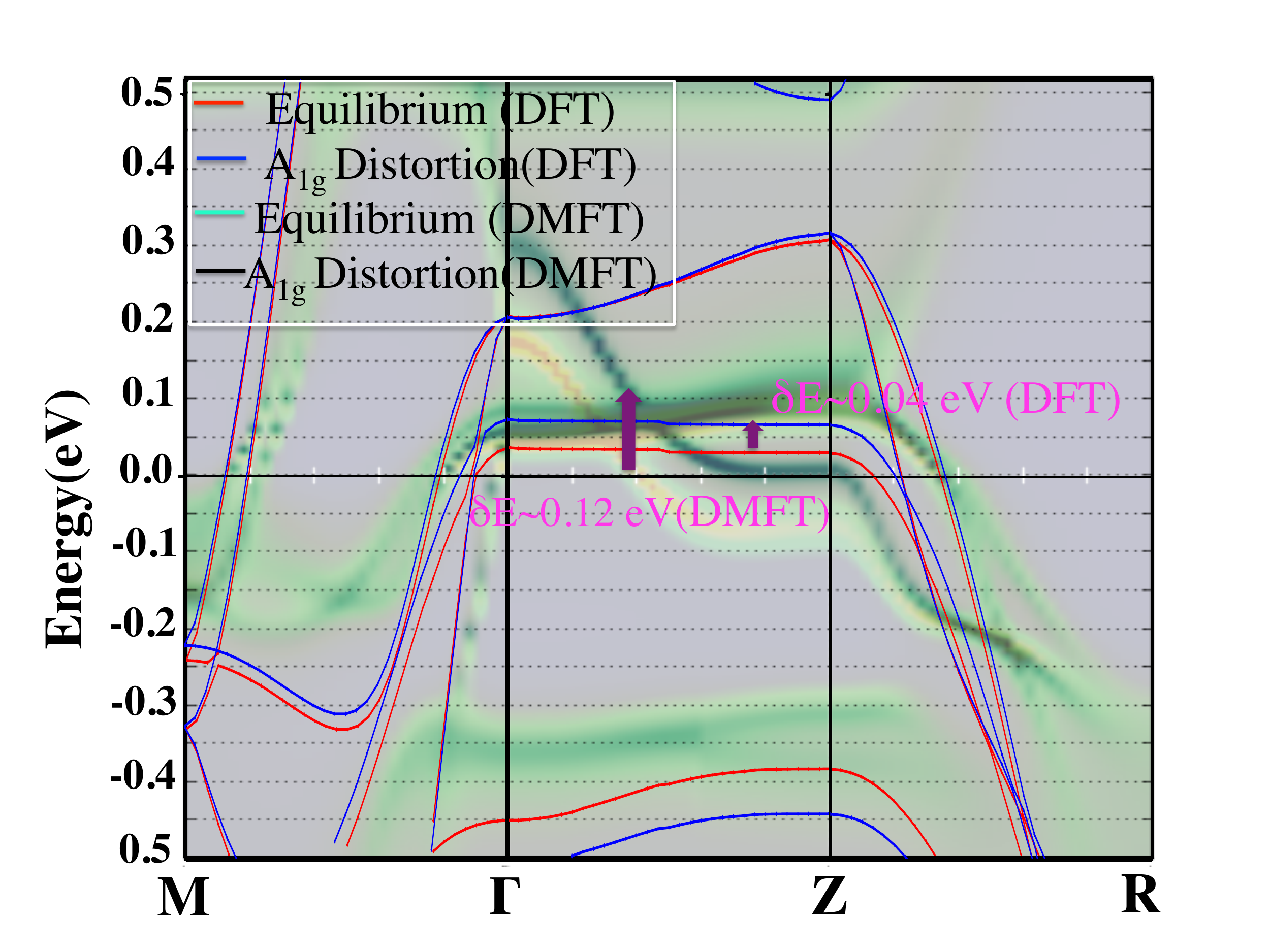}
\caption{(Color online).
(a)Maximum and Fermi surface average of deformation potentials ($\mathcal{D}$) for the A$_{1g}$ distortion computed in DFT-DMFT and DFT as a function of pressure indicates the presence of strong EPC in FeSe; inset shows deformation potential
 as a function of $E_F$ at P=0. 
(b)Demonstration of huge local electron-lattice coupling for the A$_{1g}$ distortion in our DFT-DMFT computations at P=0GPa for a selective part of the BZ; red and blue lines represent GGA bands. 
The common Fermi energy is considered for the equilibrium position and denoted by the single horizontal line for both DFT and DFT-DMFT methods.
}
\end{figure}
It is important to know how the electron correlation changes with increasing pressure. In order to investigate the degree of correlation,
we calculated  $Z_A=(1-\frac{\delta \Sigma}{\delta \omega})^{-1}_{\omega=0}$. In a Fermi-liquid it is the quasiparticle weight, which is unity for non-interacting system, and 
is much smaller than unity for strongly correlated system.
We have calculated $Z_A$ for all the Fe-d orbitals and plotted them as a function of P in Fig. 1(g).
 Though the $d_{z^2},d_{x^2-y^2}$  orbital becomes less correlated with the increase of P, the $t_{2g}$ orbitals ($d_{xz} $, $d_{yz}$, and $d_{xy}$) remain correlated 
and in particular the $d_{xy}$ orbital, which carries more magnetic moment, becomes even more correlated with increasing pressure. The predicted $Z_A$ of 0.41 for $d_{xy}$ orbital at P=0 GPa and at T=300 K
 differs from 0.287 calculated at T = 116 K in Ref.\cite{haule3} due to the use of slightly different lattice parameters and $z_{Se}$. By comparing results with Ref. \cite{haule3}, we found that among other $t_{2g}$ orbitals,
$d_{xy}$ shows the strongest temperature dependence in $Z_A$ at P=0GPa.

Calculated electron density of states, optical conductivity both along {\it ab}-plane and {\it c}-axis shows a monotonic behavior as a function of compression (details in Supplementary).
 We now turn our discussion to the electron-phonon coupling and polaron formation in FeSe. 
The coupling between electronic states and atomic displacements (the electron-phonon matrix element) and hence $\lambda$ is directly related to the shift in the energy eigenvalues
at $E_F$. We calculate this shift by calculating  $\delta E$ (details in Supplementary Information ). We estimate the average as well as the maximum value of the deformation
 potential ($\mathcal{D}=\frac{\delta E}{\delta Q}$) upon Se atom displacement ($\delta Q$) in $z_{Se}$ of a particular phonon mode
 (A$_{1g}$ distortion) for the entire FS.
In Fig. 2(a) we have plotted the deformation potential for both DFT  and 
DFT-DMFT methods.
The equation of states (pressure points on the axis) are obtained from the experiment\cite{x-ray}. 
First we notice that the average $\mathcal{D} $ increases in DFT-DMFT over that obtained in DFT for all pressure. At P=0 the average $\mathcal{D}^2 $ increases $\sim$1.5 times in DFT-DMFT;  $\mathcal{D} $ is 0.84 eV/$\AA$ in DFT-DMFT 
while 0.69 eV/$\AA$ in GGA. A similar increase was found by Boeri {\it et al.} \cite{Boeri:2010un} when magnetic softness was included in their calculations. At P=3.4 GPa the average $\mathcal{D}^2 $ increases $\sim$ 2.25 times in DFT-DMFT. 
However this is still not sufficient to obtain 37 K but not small enough to ignore as suggested earlier by Boeri {\it et al.}\cite{Boeri:2010un}. More interesting is the pressure dependence of the maximum $\mathcal{D} $. 
At ambient pressure we notice that the maximum deformation potential ($\mathcal{D}^{m}$) in DFT-DMFT is about 3 times higher than that obtained in GGA. 
One band is very sensitive to this deformation; it crosses the $E_F$ for P=0, 1.4, 2.6 and -2.0 GPa ( indicated by h1 in the Fig. 1d for P=0). 
This sensitive pocket gives a very high value of $\mathcal{D}^{m}$ for this pressure range within DFT-DMFT method, where the experimental $T_c$ is also observed to be high.
$\mathcal{D}^m$ obtained from the GGA is found to be very small at P=0.
We found that the $\mathcal{D}$ obtained from DFT-DMFT is different for different parts of the FS (details in the Supplementary Information). For example at P=0, h1 gives $\mathcal{D}^{m}$ of $\sim$ 4.4 eV/ $\AA$, while e1 gives only $\sim$ 1.5 eV/$\AA$ with DMFT method.
The largest contribution in the enhanced deformation potential is from the hole pocket (h1), located in the $\Gamma$-region. A nonuniform and anisotropic nature of the electron 
phonon coupling has also seen in the cuprates, where the average EPC was found to be one order of magnitude smaller
than the maximum \cite{Devereaux:2004gs}. This is consistent with our results.

  With increasing pressure, maximum value of the deformation potential within DFT-DMFT then decreases and remains almost unchanged after P=3.4 GPa; whereas 
the maximum $\mathcal{D}$ within DFT for this mode is insensitive to increasing pressure except at P=11 GPa, where the experimental
$T_c$ is found to decrease. At P=11 GPa, the sensitive band crosses the $E_F$ for the GGA, which leads to a high $\mathcal{D}^{m}$ for GGA. 
For pressure above 3.4 GPa, we notice that the maximum $\mathcal{D}$ in DMFT is from the electron pocket centered at $M$-point. 
$\mathcal{D}^{m}$ also strongly depends on the $E_F$. The inset of Fig. 2(a) shows the behavior of 
 maximum and average value of $\mathcal{D}$ as a function of $E_F$ at P=0 calculated with DFT-DMFT. So the movement of $E_F$ due to defect or pressure can significantly change the FS topology and hence the $\mathcal{D}$. 
 %Now to illustrate the huge deformation potential in DFT-DMFT calculation at P=0 GPa, we have plotted 
The momentum resolved spectral function $A(\omega,k)$ is shown in Fig. 2(b) for both equilibrium position and 
$A_{1g}$ distortion between the high symmetric points, where the most sensitive band crosses the $E_F$. The solid red and blue lines represent corresponding GGA bands for equilibrium position and $A_{1g}$ distortion respectively.
From the Fig. 2(b), we notice that at P=0, the shift in energy ($\delta E$) over the atomic displacement of 0.0276 $\AA$ is $\sim$ 0.12 eV in DFT-DMFT 
and $\sim$ 0.04 eV in GGA, respectively. So as reflected from Fig. 2, the $\mathcal{D}$ is about three times higher in DFT-DMFT for this particular region of the BZ.
If we notice carefully in Fig. 2(b), the shift of the bands due to $A_{1g}$ distortion is very non-uniform in DFT-DMFT; a strong deformation potential is noticed only in
 $\Gamma$ to Z region while for the other part of the BZ, deformation potential is found to be small. This leads to a strong non-uniform EPC at P=0, which is reflected
in Fig. 2(a) where maximum $\mathcal{D}$ is found to be about three times higher than the average.
 We found this similar non-uniform EPC for P=1.4, 2.6, and -2.0 GPa.

 We estimate $\lambda$ using $\mathcal{D}$ (Supplementary Information for details). While the average $\lambda$ is still small, the maximum $\lambda$ in DFT-DMFT is found to be 0.98 at P=0. At P=2.6 GPa, 
the maximum $\lambda$ reaches 1.159. We found that only certain electronic states have very strong $\lambda$ 
 while the average $\lambda$ is not strong enough to explain 37K. So a conventional electron-phonon mechanism seems unlikely. 
On the other hand, this also indicates that local EPC can be important and one can use polaron model,
where a single electron can strongly couple with the lattice and form polarons.
Formation of polaron has been experimentally found in both Fe-superconductors\cite{Liu:2013vb,NunezRegueiro:2009jp,Oyanagi:2010ep} and cuprates\cite{Zhao:1997wv}. 
The anomalous temperature dependence of the local Fe-As displacement, observed in Ref. \cite{Higashitaniguchi:2008id} indicates that 
local rather than global electron-lattice interaction is present in Fe-based superconductors and as suggested in Ref. \cite{BussmannHolder:2009fw}, 
polaron formation is responsible for the observed anomalies\cite{Higashitaniguchi:2008id}. 
Though the formation of polaron depends on a lot of factors, like the band-filling, temperature, EPC strength, phonon frequency etc, 
our results suggest to use a polaron model. We consider the electronic state corresponding to maximum $\lambda$ ($\sim$ 1) forms a polaron, 
which is a quasiparticle consisting of electron and the surrounding lattice distortion.  
Then the polaronic binding energy ($E_p$) will linearly depend \cite{Anonymous:_vFtBVCh,Anonymous:QDpFgddi} on maximum $\lambda$ and hence on square of the maximum deformation potential. 
Taking the polaronic band into account, Alexandrov and Mott \cite{Anonymous:_vFtBVCh} described that $T_c$ exponentially depends on function of $E_p$.
 Under hydrostatic pressure, we found that electronic properties change monotonically while 
only $|\mathcal{D}^{m}|^2$ (and hence $E_p$) initially grows (to 3.4 GPa)  and then drops, similarly to experimental $T_c$. This indicating that a strong local EPC plays an 
important role in Fe-based superconductors.

It is important to mention that $T_c$ was found to increase rapidly for the low pressure range (0-3 GPa) and can reach up to 27K at 1.48 GPa\cite{Mizuguchi:2008bn}.
The disagreement in the pressure dependence of experimental $T_c$ and our DFT-DMFT calculation of maximum $\mathcal{D}$ can be due to the presence of the mixed phase in low temperature crystal structure in experiment while our calculations are based on 
room temperature tetragonal (PbO-type) structure.

The behavior of the DFT-DMFT deformation potential with pressure hints that superconductivity in FeSe may have partially phonon or polaron origin and local EPI plays a very important role in superconductivity in
the unconventional superconductors. 
 Analysis of the contributions of each many-body state reveals that charge fluctuations due to correlations and charge transfer from Fe to Se are coupled to the A$_{1g}$ mode.  

Our computations predict that applied pressure significantly changes the FS around $\Gamma$-point.
We show the Fermi surface average of the deformation potential is enhanced up to 50 \% in DFT-DMFT when compared with standard
DFT, and is still not high enough to give high $T_c$ in FeSe. This reflects and confirms that a simple electron-phonon coupling mechanism seems unlikely as relevant from many experimental findings.
Calculated electronic properties show a monotonic behavior with
applied pressure.
We found a strong enhancement of the coupling between correlated electronic states along high symmetric $\Gamma$ point and $A_{1g}$ lattice distortion. 
The maximum deformation potential within GGA for this mode is almost insensitive to increasing pressure except at high pressure where the experimental $T_c$ is found to decrease.
The corresponding DFT-DMFT deformation potential behaves similarly to experimental $T_c$, which increases first and then decrease after a certain pressure. 

To explain high T$_c$ by conventional strong coupling EPI enhanced in DFT-DMFT method, these enhancements would need to be on average just as large for all modes. 
More importantly the difference (addition) between interband and intraband coupling has to be enhanced to support the $s_{\pm}$ ($s_{++}$) pairing symmetry.
Alternatively, as shown here, if only certain modes become very strongly coupled with certain electronic states, this could give sufficient coupling for polaron formation which requires only certain modes to have 
extreme coupling leading to condensation of either polaronic and/or bipolaronic states or Cooper pairs \cite{Salje:2005ts,Zhao:1997wv,Alexandrov:1994wd,Liu:2013vb}. 
For high $T_c$ cuprates, M{\"u}ller {\it et al.} showed that the normal state polaron can form Cooper pairs and can be responsible for superconductivity\cite{Zhao:1997wv}. 
Such polaronic states may involve spin as well as charge fluctuations \cite{Alexandrov:1994wd}, leading to a problem that still requires significant development for a predictive theory.
 Even if the giant correlated EPI we predict is not directly responsible for superconductivity, it should help explain some experimental observations, such as significant and unusual isotope effects \cite{Chen:2007,Dewaele:2009ur}.

The high T$_c$ superconductors so far are all rather bad metals in the normal state, have low densities of state at E$_F$, are ionic metals containing transition metal ions, and are quasi-two-dimensional and highly anisotropic, so it seems any competitive theory should explain why. Spin fluctuations may play a major role, but it seems that is not enough to explain all of these common features. The fact that they are anisotropic bad metals with low densities of states may be because this decreases the screening of the attractive interaction, partly from poorly screened Coulomb fluctuations from phonons or polarons. Correlations may enhance the EPI as we found here, and in addition lead to lower entropy from fluctuations among multiplets in the normal state leading to higher T$_c$'s.

We thank I. I. Mazin, I. I. Naumov, M. Ahart, P, Zhang, X. Chen and V. Struzhkin for helpful discussions. This research was supported as part of EFree, an Energy Frontier Research Center funded by the US Department of Energy Office of Science, Office of Basic Energy Sciences under Award DE-SC0001057. K. H acknowledges the supports from NSF DMR 0746395. Computations were performed at NERSC supercomputing facility.

\bibliography{super}
\end{document}